\documentclass[reqno]{amsproc}

\usepackage{amsmath,amssymb,amsthm,amsfonts,latexsym}
\usepackage{graphicx}
\usepackage{color}

\usepackage{txfonts}

\makeatletter
\newcommand\footnoteref[1]{\protected@xdef\@thefnmark{\ref{#1}}\@footnotemark}

  \DeclareMathOperator{\D}{d\!} 
 \DeclareMathOperator{\E}{e} \DeclareMathOperator{\I}{i}
\def\ulamek#1#2{\mbox{\normalfont$\frac{#1}{#2}$}}

\begin{document}

\title[Theory of relativistic heat polynomials \ldots]{Theory of relativistic heat polynomials and one-sided L\'{e}vy distributions }

\author{G. Dattoli}
\email{giuseppe.dattoli@enea.it}
\address{ENEA - Centro Ricerche Frascati, via E. Fermi, 45, IT 00044 Frascati (Roma), Italy}

\author{K.~G\'{o}rska}
\email{katarzyna.gorska@ifj.edu.pl}

\author{A.~Horzela}
\email{andrzej.horzela@ifj.edu.pl}
\address{H. Niewodnicza\'{n}ski Institute of Nuclear Physics, Polish Academy of Sciences, Division of Theoretical Physics, ul. Eliasza-Radzikowskiego 152, PL 31-342 Krak\'{o}w, Poland}

\author{K.~A.~Penson}
\email{penson@lptl.jussieu.fr}
\address{Sorbonne Universit\'{e}s, Universit\'{e} Pierre et Marie Curie (Paris 06), CNRS UMR 7600, Laboratoire de Physique Th\'{e}orique de la Mati\`{e}re Condens\'{e}e (LPTMC), Tour 13-5i\`{e}me \'{e}t., B.C. 121, 4 pl. Jussieu, F 75252 Paris Cedex 05, France}

\author{E.~Sabia}
\email{elio.sabia@enea.it}
\address{ENEA - Centro Ricerche Frascati, via E. Fermi, 45, IT 00044 Frascati (Roma), Italy}

\begin{abstract}
The theory of pseudo-differential operators is a powerful tool to deal with differential equations involving differential operators under the square root sign. These type of equations are pivotal elements to treat problems in anomalous diffusion and in relativistic quantum mechanics. In this paper we report on new and unsuspected links between fractional diffusion, quantum relativistic equations and particular families of polynomials, linked to the Bessel polynomials in Carlitz form, and playing the role of relativistic heat polynomials. We introduce generalizations of these polynomial families and point out their specific use for the solutions of problems of practical importance.
\end{abstract}

\maketitle

\section{Introduction}

The formalism of the evolution operator is well established in the study of partial differential evolution equations (PDEE) of Cauchy type \cite{Hadamard}. The stretching of the formalism allows the extension of the procedure to cases for which the underlying mathematical foundations are still far from being very well known \cite{Dattoli_RNC}. Such an ``abuse'' is not fully unjustified, since it leads to a wealth of old and new results, derived within a very transparent framework and with the minimum computational effort \cite{Dattoli_JPA}\footnote[1]{The correctness of the new results is usually benchmarked against different methods including numerical checks.}.

\begin{figure}[h!]
\includegraphics[scale=0.6]{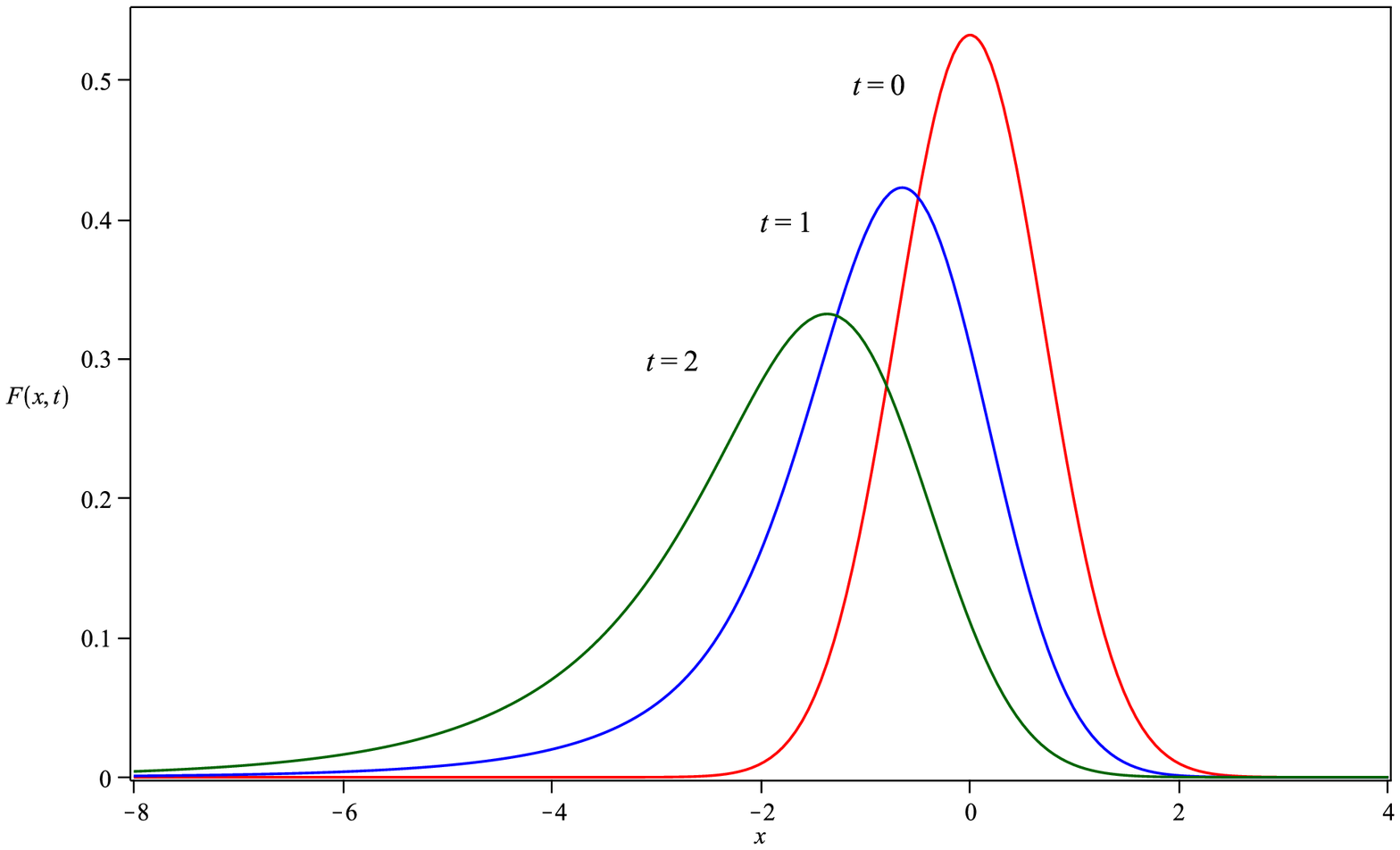}
\caption{\label{fig1} \footnotesize{The plot of $F(x,t)$ given by Eq. \eqref{eq4} for $t=0$ (blue curve), $t=1$ (red curve), and $t=2$ (green curve).}}
\end{figure}

An example of such a "temerarious" way of proceeding is the search for the solution of the PDEE
\begin{equation}\label{eq1}
\partial_t F(x, t) = \left[1-\sqrt{1-2\partial_x}\right] F(x, t), \qquad F(x, 0) = g(x),
\end{equation}
The previous equation is taylor suited for our purposes. It is indeed simple enough to be handled with advanced, but straightforward means, and is amenable for generalizations leading to interesting consequences. The relevant solution, in formal terms, is found to be
\begin{equation}\label{21/04/2016-2}
F(x, t) = \hat{U}(t) g(x), \quad \text{where} \quad \hat{U}(t) = \exp\{t[1-\sqrt{1-2\partial_x}]\}, \quad t > 0.
\end{equation}
An effective mean to evaluate the action of the evolution operator $\hat{U}(t)$ on the initial function $g(x)$ is the use of the following identity from Laplace integral transform theory \cite[Eqs. (3) and (4) for $l/k = 1/2$]{KAPenson10}, \cite[Eq. (2)]{KGorska12a}, for ${\rm Re}(p) >0$:
\begin{equation}\label{eq2}
\E^{-\sqrt{p}}=\int_0^\infty g_{1/2}(\kappa)\E^{-p\kappa} \D\kappa, \qquad g_{1/2}(\kappa) = \frac{\exp[-1/(4\kappa)]}{2\sqrt{\pi}\kappa^{3/2}},
\end{equation}
where $g_{1/2}(\kappa)$ is the L\'{e}vy-Smirnov (L-S) distribution and it is a canonical example of one-sided stable distributions. For various representations and transformation properties of $g_{l/k}$ for general integer $l$ and $k$ consult \cite{KAPenson10, KGorska12a, GP, GHPD, PG}. According to Eq. \eqref{eq2} the evolution operator can be cast in the form
\begin{equation}\label{21/04/2016-4}
\hat{U}(t)=\E^t \int_0^\infty g_{1/2}(\kappa) \E^{-\kappa t^2} \E^{2 \kappa t^2 \partial_x} \D\kappa
\end{equation}
an therefore
\begin{equation}\label{eq3}
F(x, t) = \E^t \int_0^\infty g_{1/2}(\kappa) \E^{-\kappa t^2} g(x+2\kappa t^2) \D\kappa. 
\end{equation}
If, for example, $g(x) = e^{-x^2}/\sqrt{\pi}$ we get
\begin{equation}\label{eq4}
F(x, t) = \varphi(x, t) \frac{\E^{-x^2}}{\sqrt{\pi}}, \quad \varphi(x, t) = \E^{t} \int_0^\infty g_{1/2}(\kappa) \E^{-\kappa t^2} \E^{-4\kappa t^{2}(x+\kappa t^{2})} \D\kappa.
\end{equation}
The evolution of an initial Gaussian distribution ruled by Eq. \eqref{eq1} is provided in Fig. \ref{fig1}.

As a further example of initial condition we consider the case of a monomial, namely $g(x) = x^n$, for which Eq. \eqref{eq3} yields 
\begin{align}\label{eq5}
\begin{split}
F(x, t) &= \E^{t}\int_0^\infty g_{1/2}(\kappa) \E^{-\kappa t^2} (x+2\kappa t^2)^n \D\kappa = \sum_{k=0}^n \binom{n}{k} x^{n-k} B_{k}(t), \\
B_k(t) & = \E^t \int_0^\infty g_{1/2}(\eta) \E^{-\eta t^2} (2\eta t^2)^k \D\eta.
\end{split}
\end{align}
It turns out that $B_{k}(t)$ are polynomials of order $k$. The nature of $B_{k}(t)$ polynomials can be straightforwardly understood. We note that the relevant exponential generating function can be obtained from Eq. \eqref{eq2} and the second of Eqs. \eqref{eq5}: 
\begin{equation}\label{21/04/2016-7}
\sum_{n=0}^{\infty} \frac{\xi^n}{n!} B_{n}(t) = \E^t \int_0^\infty g_{1/2}(\kappa) \E^{-\kappa t^2} \E^{-2\kappa\xi t^2}\D\kappa = \exp[t(1-\sqrt{1-2\xi})].
\end{equation}
$B_{n}(t)$ are therefore  recognized as Bessel polynomials (in Carlitz form) \cite{LCarlitz57} specified by the equation
\begin{equation}\label{21/04/2016-8}
B_{n}(t) = \sum_{k=1}^{n} \frac{(2n-k-1)!}{(k-1)! (n-k)!}\, \frac{t^k}{2^{n-k}}.
\end{equation}

In this introductory section we have shown that evolution equations of the type \eqref{eq1} involving a pseudo-differential operator naturally leads to Carlitz type polynomials. We will see in the following that this is quite a general property. We will, in particular prove that PDEE, ruled by square root differential operators, can be  solved using integral transform involving the L-S distributions, in turn naturally linked to the Bessel-Carlitz polynomials.   

\section{The relativistic heat polynomials}
Before proceeding with the specific topic of this section we want just to underscore that using the generating function of Bessel polynomials, we can express the evolution operator associated with Eq. \eqref{eq1} as the following expansion
\begin{equation}\label{21/04/2016-9}
\E^{t(1-\sqrt{1-\partial_x})} = \sum_{n=0}^{\infty} \frac{B_n(t)}{2^n n!} \partial_x^n.
\end{equation}
We note that, by keeping the second order in the square root expansion in $\partial_x$, we end up with a diffusion type equation. The ordinary heat equation has naturally lead to the introduction of the heat polynomials \cite{Dattoli_RNC} (see below), recognized as special cases of the Hermite-Kamp\'{e} de F\'{e}riet family (H-KdF) \cite{Dattoli_RNC, Dattoli_YuABrychkov08}. The higher order heat equations have also been shown to be associated with higher order Hermite-like forms \cite{KAPenson03}. Following the same stream, we will introduce below a new family of polynomials as a natural consequence of Eqs. \eqref{eq1} and \eqref{eq5}.

To this aim we use  the  umbral notation
\begin{equation}\label{21/04/2016-10}
\sum_{n=0}^{\infty} \frac{B_n(t)}{2^n n!} \partial_x^n = \exp\left(\frac{\hat{b}}{2}\partial_x\right), \qquad \hat{b}^n = B_n(t)
\end{equation}
allowing the restyling of the evolution operator in an exponential like form. The relevant action on an initial function $g(x)$ yields
\begin{equation}\label{eq6}
\E^{\frac{\hat{b}}{2}\partial_x}g(x)=g\left(x+\frac{\hat{b}}{2}\right).
\end{equation}
For $g(x)=x^n$ it allows to introduce the polynomials ${_R N_n}(x, t)$ defined as follows, using Eq.~\eqref{21/04/2016-10}:
\begin{equation}\label{21/04/2016-12}
\E^{\frac{\hat{b}}{2}\partial_x}x^n ={_R N_n}(x, t)=\left(x+\frac{\hat{b}}{2}\right)^n =\sum_{s=0}^n \binom{n}{s} x^{n-s} \frac{B_{s}(t)}{2^s}.
\end{equation}
They will be called henceforth the relativistic Newton polynomials (RNP) and are easily recognized to belong to the Sheffer family \cite{BGHP, Dattoli_YuABrychkov08}. The exponenial generating function is indeed
\begin{equation}\label{eq7}
\sum_{n=0}^{\infty}\frac{u^n}{ n!} {_R N_n}(x, t)=\E^{u\left(x+\frac{\hat{b}}{2}\right)}=\E^{ux}\E^{t(1-\sqrt{1-u})},
\end{equation}
which follows from Eq. \eqref{21/04/2016-9} upon replacing $\partial_{x}\to u$, and therefore it is also checked that
\begin{equation}\label{eq8}
\partial_x\ [{_R N_n}(x, t)]= n\cdot   {_R N_{n-1}}(x, t)
\end{equation}
The Bessel-Carlitz polynomials satisfy the recurrence
\begin{equation}\label{21/04/2016-15-1}
B_n^{''}(t)-2B_n^{'}(t)=-2nB_{n-1}(t)
\end{equation}
and by merging  Eqs. \eqref{eq7} and \eqref{eq8}, we end up with 
\begin{equation}\label{eq9}
-\partial_t^2 {_R N_n}(x, t) + 2 \partial_t\, {_R N_n}(x, t) = \partial_x\, {_R N_n}(x, t).
\end{equation}
apparently different from our starting Eq. \eqref{eq1}. Both Eqs. \eqref{eq1} and \eqref{eq9} will be reconciled in the concluding section. 

\begin{figure}[h!]
\includegraphics[scale=0.6]{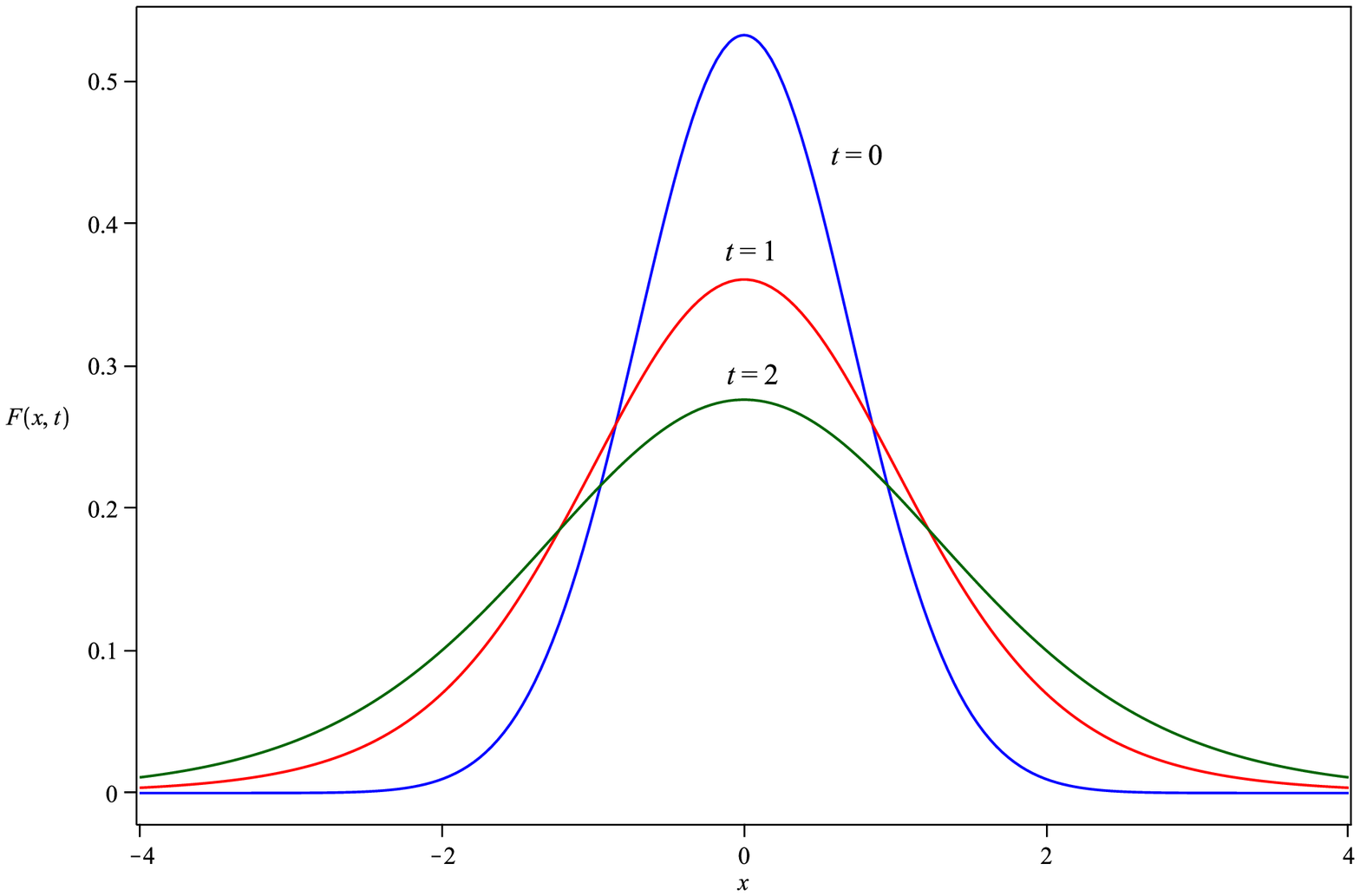}
\caption{\label{fig2} \footnotesize{The plot of $F(x, t)$ given by Eq. \eqref{eq11} for $t=0$ (blue curve), $t=1$ (red curve), and $t=2$ (green curve).}}
\end{figure}

The present authors have introduced the relativistic heat equation in the form \cite{Dattoli_JPA}
\begin{equation}\label{eq10}
\partial_t F(x, t) = (1-\sqrt{1-\partial_x^2})F(x, t), \quad F(x,0) = g(x).
\end{equation}
The elimination of the square root is obtained as indicated in the previous section by the use of a Laplace transform, thus reducing the problem to an ordinary heat equation, whose solution is obtained in terms of the Gauss-Weierstrass transform, namely
\begin{align}\label{21/04/2016-14}
\begin{split}
F(x, t) & = \E^t\int_0^\infty g_{1/2}(\eta)\E^{-\eta t^2} \left[\E^{\eta t^2 \partial_x^2} g(x)\right] \D\eta \\
& = \E^t \int_0^\infty g_{1/2}(\eta) \E^{-\eta t^2} \left[\frac{1}{2t\sqrt{\pi\eta}} \int_{-\infty}^{\infty} \E^{-\frac{(x-\sigma)^2}{4\eta t^2}} g(\sigma) \D\sigma\right]\D\eta.
\end{split}
\end{align}
Using once more a Gaussian as initial condition and applying the Glaisher formula  \cite{Dattoli_AMC} we find 
\begin{equation}\label{eq11}
F(x, t) = \frac{\E^t}{\sqrt{\pi}} \int_0^\infty g_{1/2}(\eta) \E^{-\eta t^2} \frac{\exp\left(-\frac{x^2}{1+4\eta t^2}\right)}{\sqrt{1+4\eta t^2}} \D\eta,
\end{equation}
which is presented in Fig. \ref{fig2} for several values of $t$. 

On the other side by keeping as initial condition a monomial $x^{n}$ we find
\begin{equation}\label{21/04/2016-17}
F(x, t) = \E^t \int_0^\infty g_{1/2}(\eta)\E^{-\eta t^2} \left(\E^{\eta t^2 \partial_x^2} x^n\right) \D\eta = \E^t \int_0^\infty g_{1/2}(\eta)\E^{-\eta t^2} H_n(x, \eta t^2) \D\eta.
\end{equation}
The previous result has been obtained by taking into account that the two-variable Hermite-Kamp\'{e} de F\'{e}riet (the two-variable H-KdF) polynomials $H_n(x, y)$ with the generating function \cite{Dattoli_RNC}:
\begin{equation}\label{eq12}
\sum_{n=0}^{\infty} \frac{t^n}{n!} H_n(x, y) = \E^{xt+y t^2},
\end{equation}
are a natural solution of the ordinary diffusion equation
\begin{equation}\label{21/04/2016-18b}
\E^{y\partial_x^2} x^n = H_n(x, y), \quad H_n(x, y) = n! \sum_{r=0}^{\lfloor n/2 \rfloor}\frac{x^{n-2r} y^r}{(n-2r)! r!}.
\end{equation}
For this reason the H-KdF polynomials are sometimes called heat polynomials. The polynomials $H_{n}(x, y)$ are related to conventional Hermite polynomials $H_{n}(z)$ through $H_{n}(x, y) = (-\I)^{n} y^{n/2} H_{n}(\I x/2\sqrt{y})$, see \cite{Dattoli_RNC}. We will accordingly introduce the relativistic heat polynomials (RHP) defined by the operational identity
\begin{equation}\label{eq13}
{_R H_n}(x, t) = \E^{t(1-\sqrt{1-\partial_x^2})} x^n.
\end{equation}
By following the same considerations as exploited for RNP, we find for the RHP the result
\begin{equation}\label{21/04/2016-20}
{_R H_n}(x, t) = \sum_{r=0}^{\infty}\frac{B_r(t)}{2^r r!} \partial_{x}^{2r} x^n = n! \sum_{r=0}^{\lfloor n/2 \rfloor} \frac{x^{n-2r}}{(n-2r)! r!}\,\frac{B_r(t)}{2^r} = H_n\left(x,\frac{\hat{b}}{2}\right).
\end{equation}
According to Eq. \eqref{eq6} we can also conclude that the RNP satisfy the PDEE
\begin{equation}\label{eq14}
-\partial_t^2 {_R H_n}(x, t) + 2\partial_t\, {_R H_n}(x, t) = \partial_x^2\, {_R H_n}(x, t),
\end{equation}
which enforces our assumption that the equations we are dealing with are "relativistic". Eq. \eqref{eq14} can be recognized as the relativistic heat equation proposed to solve the problems associated with infinite speed of propagation of heat inside solid and it is indeed the one dimensional telegrapher equation (provided that $t\to -t$, $x\to \I x$).  The RHP associated with the telegrapher equation \cite{Cattaneo} are therefore given by ${_R H_n}(\I x, -t)$.

We have so far been able to reconcile the theory of L\'{e}vy-stable distributions with heat type relativistic equation and we have introduced the RHP are Bessel polynomials-based H-KdF polynomials.

\begin{figure}[h!]
\includegraphics[scale=0.6]{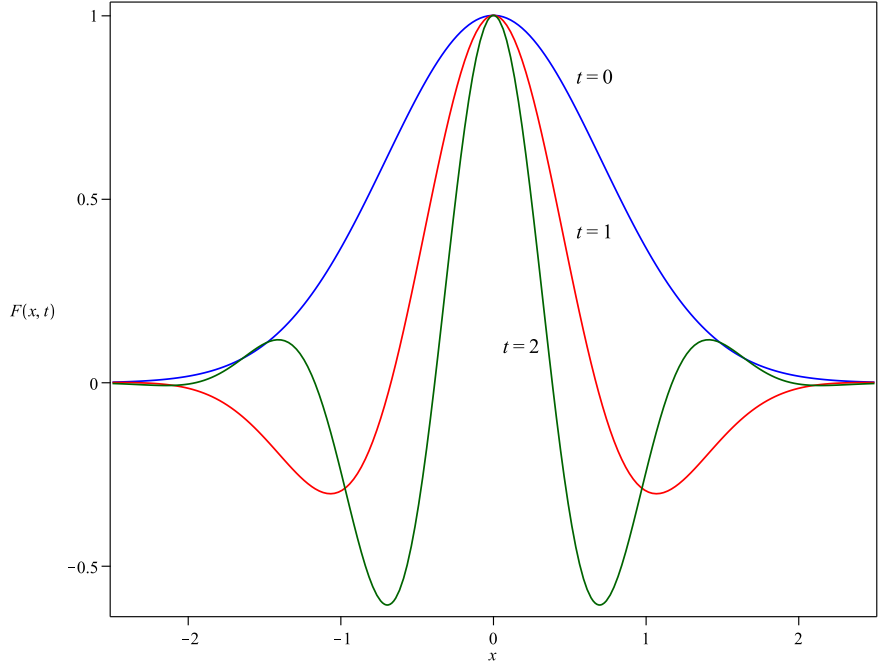}
\caption{\label{fig3} \footnotesize{The plot of $F(x, t)/F(0, t)$, where $F(x, t)$ is given by Eq. \eqref{eq15} for $t=0$ (blue curve), $t=1$ (red curve), and $t=2$ (green curve).}}
\end{figure}

It should however be noted that Eq. \eqref{eq14} is more general that the fractional Eq. \eqref{eq10}; it is indeed a hyperbolic type equation
\begin{equation}\label{21/04/2016-22}
\partial_t^2 F(x, t) - 2\partial_t F(x, t) + \partial_x^2 F(x, t) = 0
\end{equation}
 with the initial conditions
\begin{equation}\label{1/12-1}
F(x, 0) = g(x), \quad \text{and} \quad \partial_t F(x, t)\Big\vert_{t=0} = s(x).
\end{equation}
The general solution can accordingly be written as
\begin{equation}\label{21/04/2016-23}
F(x, t) = \E^t \left[\E^{t\sqrt{1-\partial_x^2}} C_1(x) + \E^{-t\sqrt{1-\partial_x^2}} C_2(x)\right],
\end{equation}
where the functions $C_1(x)$ and $C_2(x)$ are specified through the initial conditions and therefore in operational form we find
\begin{equation}\label{21/04/2016-24}
F(x, t) = \E^t \left\{\cosh(t\sqrt{1-\partial_x^2}) g(x) + \frac{\sinh(t\sqrt{1-\partial_x^2})}{\sqrt{1-\partial_x^2}}[s(x) - g(x)]\right\}.
\end{equation}
The assumption that the norm of the function be a conserved quantity implies that $s(x)=0$. Examples of solutions for the gaussian initial condition are given in Fig. \ref{fig3}. More precisely, in Fig. \ref{fig3} are presented the plots obtained from the formula below:
\begin{equation}\label{eq15}
F(x, t) =\frac{\E^{t}}{2\pi} \int_{-\infty}^{\infty} \E^{\I k x} \left[\cosh(t\sqrt{1+k^{2}}) - \frac{\sinh(t\sqrt{1+k^{2}})}{\sqrt{1+k^{2}}}\right] \E^{-k^{2/4}} \D k,
\end{equation}
which includes the Fourier transform of the gaussian initial condition and $s(x) = 0$. We do not further dwell on the physical implications of relativistic equation and leave the problem for future investigations.

\section{RHP as a pseudo-orthogonal basis}

According to the previous description the RHP are obtained from the operational condition in Eq. \eqref{eq13}. We have stressed their similarity with ordinary heat polynomials (the two-variable H-KdF polynomials) and therefore we may also ask what are the conditions to expand a given function in terms of RHP. To this aim we use the technique proposed in Refs. \cite{Dattoli_AMC,Dattoli_ITSF}, where the orthogonality properties of Hermite and Laguerre type polynomials has been studied by just starting from the relevant operational definitions.

We will therefore assume that if a function can be expanded in terms of RHP the following identity holds 
\begin{equation}\label{21/04/2016-25}
f(x) = \sum_{r=0}^\infty c_r\, {_R H_r}(x, -|y|)
\end{equation}
(note that within the present context $y$ is a constant and therefore we are not dealing with two variables). According to Eq. \eqref{eq13} and assuming that it can also be inverted, we find
\begin{equation}\label{eq16}
\E^{y(1-\sqrt{1-\partial_x^2})} f(x) = \sum_{n=0}^\infty c_n(y) x^n.
\end{equation}
The use of the transforms reported in the introductory remarks allows to cast Eq. \eqref{eq16} in the form
\begin{equation}\label{21/04/2016-27}
\E^{y} \int_0^\infty g_{1/2}(\eta) \E^{-\eta y^2} \E^{|y|^2 \eta \partial_x^2} f(x)\D\eta = \sum_{n=0}^\infty c_n(y) x^n.
\end{equation}
The application of the Gauss-Weierstrass transform allows the evaluation of the action on $f(x)$ of the integral with the second derivative and therefore we find
\begin{equation}\label{21/04/2016-28}
\left[\E^y \int_0^\infty g_{1/2}(\eta) \E^{-\eta y^2} \E^{\eta\, |y|\, \partial_x^2}\D\eta\right] f(x)  = \frac{\E^y}{2|y| \sqrt{\pi}} \int_0^\infty g_{1/2}(\eta) \frac{\E^{-\eta y^2}}{\sqrt{\eta}} \left[\int_{-\infty}^{\infty} \E^{-\frac{(x-\xi)^2}{4y^2\eta}} f(\xi) \D\xi\right] \D\eta. 
\end{equation}
Finally, according to Eq. \eqref{eq12}, we end up with 
\begin{align}\label{eq17}
\begin{split}
\frac{\E^y}{2|y|\sqrt{\pi}} \int_0^\infty g_{1/2}(\eta) \frac{\E^{-\eta y^2}}{\sqrt{\eta}} \left[\int_{-\infty}^{\infty}\E^{-\frac{(x-\xi)^2}{4y^2 \eta}} f(\xi) \D\xi\right] \D\eta \qquad\qquad\qquad\qquad\qquad\qquad\qquad \\
= \sum_{n=0}^{\infty} \frac{x^n}{n!}\left\{\frac{\E^y}{2|y|\sqrt{\pi}} \int_0^\infty g_{1/2}(\eta) \frac{\E^{-\eta y^2}}{\sqrt{\pi}}\left[\int_{-\infty}^{\infty} H_n(\ulamek{\xi}{4 y^2 \eta}, -\ulamek{1}{4 y^2 \eta})\E^{-\frac{\xi^2}{4 y^2 \eta}}\D\xi\right]\D\eta\right\}.
\end{split}
\end{align}
The coefficients $c_n$ are  obtained by confronting the coefficients of the same power of $x$, thus getting
\begin{align}\label{eq18}
\begin{split}
c_n(y) & = \frac{\E^y}{2|y|\sqrt{\pi}} \int_0^\infty g_{1/2}(\eta) \frac{\E^{-\eta y^2}}{\sqrt{\eta}} \chi_n(\eta) \D\eta,\\
\chi_n(\eta) & = \frac{1}{n!} \int_{-\infty}^{\infty} H_n(\ulamek{\xi}{4 y^2 \eta}, -\ulamek{1}{4 y^2 \eta}) \E^{-\frac{\xi^2}{4y^2 \eta}} f(\xi) \D\xi.
\end{split}
\end{align}
Accordingly the expansion in terms of RHP is possible whenever the integral in Eq. \eqref{eq18} converge.

It is now worth noting that the coefficients $\chi_n(\eta)$ are those of the expansion of the function $f(x)$ in Hermite polynomials. Even though we cannot explicitly state any orthogonal properties of RHP, we can say that the relevant pseudo-orthogonality is, in some sense, a by-product of that of the ordinary Hermite  and that it is associated with a convolution with the  L\'{e}vy distribution function $g_{1/2}(\eta)$.

\section{Generalization of Bessel-Carlitz polynomials and the generalization of RHP polynomials}

It is almost natural to consider extensions of Eq. \eqref{eq1} to forms including second derivatives as e.g.:
\begin{equation}\label{eq19}
\partial_t F(x, t) = [1-\sqrt{1 - (\alpha \partial_x^2 + \beta \partial_x)}] F(x, t), \quad F(x, 0) = g(x),
\end{equation}
with $\alpha, \beta > 0$. In this case the relevant solution, for a Gaussian as initial condition, can be written as
\begin{equation}\label{21/04/2016-32}
F(x, t) = \frac{\E^t}{\sqrt{\pi}} \int_0^\infty g_{1/2}(\eta) \E^{-\eta t^2} \frac{1}{\sqrt{1+4\alpha\eta t^2}} \exp\left[-\frac{(x+\beta\eta t^2)^2}{1+4\eta t^2}\right] \D\eta,
\end{equation}
which, according to the previous discussion is almost trivial; it is however fairly interesting  to investigate whether a generalized family of Bessel polynomials emerges from such a procedure, allowing e.g. the formal solution of Eq. \eqref{eq19} as 
\begin{equation}\label{eq20}
F(x, t) = \E^{t[1-\sqrt{1-(\alpha\partial_x^2 + \beta\partial_x)}]} g(x) = \sum_{n=0}^{\infty} \frac{\tilde{B}_n(\alpha, \beta; t)}{n!} \partial_x^n g(x).
\end{equation}
The assumption \eqref{eq20} implicitly yields the form of the exponential generating function of the generalized Bessel polynomials, therefore we find
\begin{equation}\label{21/04/2016-34}
\sum_{n=0}^{\infty} \frac{\lambda^n}{n!} \tilde{B}_n(\alpha, \beta; \sigma) = \E^\sigma \int_0^\infty g_{1/2}(\xi) \E^{-\xi\sigma^2[1-(1+\alpha \lambda^2 + \beta \lambda)]}\D\xi
\end{equation}
After using the H-KdF polynomials generating function we obtain from Eq. \eqref{eq17}  the following integral transform defining the generalized Bessel polynomials
\begin{equation}\label{21/04/2016-35}
\tilde{B}_n(\alpha, \beta; \sigma) = \E^{\sigma} \int_0^\infty g_{1/2}(\xi) \E^{-\xi\sigma^2} H_{n}(\beta\xi\sigma^2, \alpha\xi\sigma^2)  \D\xi.
\end{equation}
It is useful to note that, according to the use of the following identity
\begin{equation}\label{21/04/2016-36}
\int_0^\infty\!\! g_{1/2}(\xi) \E^{-\xi\sigma^2} \xi^n \D\xi = (-1)^n \partial_{\sigma^2}^n \int_0^\infty\!\! g_{1/2}(\xi) \E^{-\xi\sigma^2} \D\xi = (-1)^n \partial_{\sigma^2}^n \E^{-\sigma}
\end{equation}
we can derive the following Rodrigues-like formulae for the Bessel type polynomials
\begin{align}\label{21/04/2016-37}
\begin{split}
B_n(\sigma) & = (-1)^n \sigma^{2n} \E^\sigma (\ulamek{1}{\sigma}\partial_\sigma)^n \E^{-\sigma} \\
\tilde{B}_n(\alpha, \beta; \sigma) & = \sigma^{2n} \E^{\sigma} H_n(\beta\partial_{\sigma^2}, \alpha\partial_{\sigma^2})\E^{-\sigma}.
\end{split}
\end{align}
It is now evident that these generalizations can be pursued \textit{ad libitum} and in the concluding section we will give a short account of possible developments in this direction. 

\section{Concluding Remarks}

In Sec. 2 we have stressed that the RNP satisfy Eq. \eqref{eq9}, which can be recast in the form
\begin{equation}\label{eq21}
(\partial_t-1)^2  { }{_R N_n}(x, t)=(1-\partial_x){ }{_R N_n}(x, t).
\end{equation}
It is evident that subset of its general solution can be written as
\begin{equation}\label{17/06/2016-37}
{_R N_n}(x, t)=\E^t\left[\E^{t\sqrt{1-\partial_x}} c_1(x)+\E^{-t\sqrt{1-\partial_x}} c_2(x)\right].
\end{equation}
For $c_1(x)=0$, $c_2(x)=x^n$  and this reconciles the apparent inconsistency with Eq. \eqref{eq1} from which originated their definition.

Let us now consider the equation for the RHP which will be rewritten as
\begin{equation}\label{eq22}
(\partial_t-1)^2  {_{R} H_{n}(x, t)} = (1-\partial_x^2) {_{R} H_{n}(x, t)}.
\end{equation}
Unlike Eq. \eqref{eq21} it contains a second derivative in $x$ and we ask whether it can be transformed into an equation containing first derivative only by means of the Dirac type factorization \cite{Dattoli_JPA} To this end we write
\begin{equation}\label{17/06/2016-39}
(\partial_t-\hat1)\underline{\Phi}=\hat\sigma_1\underline{\Phi}+\I\hat\sigma_2\partial_x\underline{\Phi}, \quad \underline{\Phi}=\binom{\phi_1}{\phi_2},
\end{equation}
where $\hat{\sigma}_{i}$, $i=1, 2, 3$  are the Pauli matrices satisfying the properties
\begin{equation}\label{17/06/2016-40}
\hat\sigma_1^2=\hat\sigma_2^2=\hat\sigma_3^2=\hat1,\qquad \left[\hat\sigma_l, \hat\sigma_m\right]=\I\varepsilon_{l m n}\hat\sigma_n,\qquad \hat\sigma_l\hat\sigma_m+\hat\sigma_m\hat\sigma_l=0.
\end{equation}
$\hat1$ denotes the unit matrix and $\varepsilon_{l m n}$ is the Levi-Civita tensor. The price we have paid is that we have reduced the previous equation to a two-component form, namely
\begin{equation}\label{17/06/2016-41}
\partial_t\binom{\phi_1}{\phi_2}=\left(\begin{matrix} 1 & \frac{1}{2}(1+\partial_x) \\ \frac{1}{2}(1-\partial_x) & 1 \end{matrix}\right)\binom{\phi_1}{\phi_2}.
\end{equation}
The consequence of the use of such a factorization is that an equation of the type \eqref{eq22}, already exploited to treat relativistic heat conduction \cite{Cattaneo, Zhukowsky}, is transformed into a (two component) Dirac-like equation. We have underscored this result, even though slightly out of the scope of the paper, because it is interesting and indicates the wealth of the possible implications for the problem we have studied. We believe however that such an indication should be more thoroughly pursued.

Let us now outline further results which can be obtained by stretching further the formalism we have proposed.

One of the crucial elements of our discussion has been the use of Laplace transform involving L\'{e}vy stable distribution, exploited to overcome the exponentiation of a pseudo-differential operators. The L-S distribution $g_{1/2}(\eta)$  is a particular case of more general forms of two-sided distributions $g_{l/k}(\eta)$, $l,k\in\mathbb{N}$,  $l<k$ \cite{KAPenson10, KGorska12a, GP, PG}, having the property
\begin{equation}\label{eq23}
e^{-p^{l/k}} = \int_0^\infty e^{-p u} g_{l/k}(u) du, \quad 0 < l/k < 1,
\end{equation}
It is therefore evident that the use of the previous equation can produce the solution of the PDEE 
\begin{equation}\label{eq24}
\partial_{t} F(x, t) = [1-(1-\partial_{x}^{\mu})^{l/k}] F(x, t), \quad F(x, 0) = g(x), \quad \mu,l, k \in \mathbb{N}, \quad
l<k,
\end{equation}
in terms of L\'{e}vy convolutions. Furthermore it allows one to introduce the associated Bessel-Carlitz polynomials, according to the extension of previous procedure. The formal solution of Eqs. \eqref{eq24} can be written as the action of evolution operator $\hat{U}_{l, k}^{\mu}(t)$ acting on the initial condition $g(x)$:
\begin{equation}\label{6/05/2016-1}
F(x, t) = \hat{U}_{l, k}^{\mu}(t) g(x), \qquad \hat{U}_{l, k}^{\mu}(t) = \E^{t[1-(1-\partial_{x}^{\mu})^{l/k}]}.
\end{equation}
The evolution operator is related to the polynomials $B^{(l, k)}_{n}(t)$ through the exponential generating function as follows
\begin{equation}\label{20/12/2015-3}
\sum_{n=0}^{\infty} \frac{\lambda^{n}}{n!} B^{(l, k)}_{n}(\sigma) = \exp\{\sigma[1 - (1 - \ulamek{k}{l}\lambda)^{l/k}]\}.
\end{equation}
The use of Eq. \eqref{eq23} finally allows the following identification:
\begin{equation}\label{eq25}
B^{(l, k)}_{n}(\sigma) = \left(\frac{k}{l} \sigma^{k/l} \right)^{n} \E^{\sigma} \int_{0}^{\infty} \E^{-u\sigma^{k/l}} g_{l/k}(u) u^n du.
\end{equation}
Observe that $B_{n}(\sigma)$ given in Eq. \eqref{eq5} is the special case of Eq. \eqref{eq25} with $l=1$ and $k=2$. In general form Eq. \eqref{eq25}, the polynomials $B^{(l, k)}_{n}(\sigma)$ are given through properly regularized (via the exponential) moments at one-sided L\'{e}vy stable distribution functions. Representing $u^n \exp(-u\sigma^{k/l})$ as the $n$th derivative, i.e. $u^n \exp(-u\sigma^{k/l}) = (-1)^n \ulamek{d^n}{d a^n} \exp(-u a)\vert_{a=\sigma^{k/l}}$, and using Eq. \eqref{eq23} we arrive at the Rodrigues-type formula:
\begin{equation}\label{20/10/2015-5b} 
B^{(l, k)}_{n}(\sigma) 
= ( -\sigma^{\,k/l})^{n} \E^{\sigma} \left(\frac{1}{\sigma^{k/l - 1}} \frac{d}{d\sigma}\right)^n \E^{-\sigma}.
\end{equation}

In this paper we have presented new results concerning the solution of PDEE containing pseudo-differential operators \cite{Lam, BDQ}, revealing previously unnoticed links to special polynomials, special functions and L\'{e}vy type transforms. We have just outlined the lines along which the present research can be developed, and many of these topics will be explored in forthcoming investigations.

\section*{Acknowledgments}
K. G., A. H. and K. A. P. were supported by the PAN-CNRS program for French-Polish collaboration and the BGF scholarship founded by French Embassy in Warsaw, Poland. Moreover, K. G. thanks for support from MNiSW (Warsaw, Poland), "Iuventus Plus 2015-2016", program no IP2014 013073.


\begin{thebibliography}{10}

\bibitem{Hadamard} J. Hadamard, ``Lectures on Cauchy's problem in linear partial differential equations", (Dover Publications, London, 1952).

\bibitem{Dattoli_RNC} G. Dattoli, P. L. Ottaviani, A. Torre, and L. Vazquez, \textit{Evolution operator equations: Integration with algebraic and finitedifference methods. Applications to physical problems in classical and quantum mechanics and quantum field theory}, La Rivista del Nuovo Cimento \textbf{20}(2), 3 (1997).

\bibitem{Dattoli_JPA} G. Dattoli, E. Sabia, K. G\'{o}rska, A. Horzela, and K. A. Penson, \textit{Relativistic wave equations: an operational approach}, J. Phys. A: Math. Theor. \textbf{48}, 125203 (2015).

\bibitem{KAPenson10} K. A. Penson and K. G\'{o}rska, \textit{Exact and explicit probability densities for one-sided L\'{e}vy stable distributions}, Phys. Rev. Lett. \textbf{105}, 210604 (2010).

\bibitem{KGorska12a} K. G\'{o}rska and K. A. Penson, \textit{L\'{e}vy stable distributions via associated integral transform}, J. Math. Phys. \textbf{53}, 053302 (2012).

\bibitem{GP} K. G\'{o}rska and K. A. Penson, \textit{Exact and explicit evaluation of Br\'{e}zin-Hikami kernels}, Nucl. Phys. B \textbf{872}, 333 (2013).

\bibitem{GHPD} K. G\'{o}rska, A. Horzela, K. A. Penson, and G. Dattoli, \textit{The Higher-Order Heat-Type Equations via signed L\'{e}vy stable and generalized Airy functions}, J. Phys. A: Math. Theor. \textbf{46}, 425001 (2013).

\bibitem{PG} K. A. Penson and K. G\'{o}rska, \textit{On the properties of Laplace transform originating from one-sided L\'{e}vy stable laws}, J. Phys. A: Math.Theor. \textbf{49}, 065201 (2016).

\bibitem{LCarlitz57} L. Carlitz, \textit{A note on the Bessel polynomials}, Duke Math. J. \textbf{24}, 151 (1957).

\bibitem{BGHP} P. B{\l}asiak, G. Dattoli, A. Horzela, and K. A. Penson, \textit{Representations of monomiality principle with Sheffer-type polynomials and boson normal ordering}, Phys. Lett. A \textbf{352}, 7 (2006).


\bibitem{Dattoli_YuABrychkov08} G. Dattoli, \textit{Hermite-Bessel and Laguerre-Bessel functions: a by-product of the monomiality principle}, in: D. Cocolicchio, G. Dattoli, H.M. Srivastava, (Eds.) ``Advanced special functions and applications" (Melfi, 1999), in: Proc. Melfi Sch. Adv. Top. Math., vol. 1, Rome, 2000, pp. 147--164; 

\bibitem{KAPenson03} K. G\'{o}rska, A. Horzela, K. A. Penson, and G. Dattoli, \textit{The Higher-Order Heat-Type Equations via signed L\'{e}vy stable and generalized Airy functions},  J. Phys. A: Math. Theor. \textbf{46}, 425001 (2013) .


\bibitem{Dattoli_AMC} G. Dattoli, B. Germano, P. E. Ricci, \textit{Comments on monomiality, ordinary polynomials and associated bi-orthogonal functions}, Appl. Math. Comput. \textbf{154}, 219 (2004).

\bibitem{Dattoli_ITSF} G. Dattoli, S. Khan, and P. E. Ricci, \textit{On Crofton-Glaisher type relations and derivation of generating functions for Hermite polynomials including the multi-index case}, Integr. Transf. Spec. Fun. \textbf{19}, 431 (2008) and references therein.

\bibitem{Cattaneo} C. R. Cattaneo, \textit{Sur une forme de l'\'{e}quation de la chaleur \'{e}liminant le paradoxe d'une propagation instantan\'{e}e}, C. R. Ac. Sci. \textbf{247}(4), 431 (1958).

\bibitem{Zhukowsky} K. V. Zhukowsky, \textit{Operational method of solution of linear non-integer ordinary and partial differential equations}, Springer Plus \textbf{5}, 119 (2016).

\bibitem{Lam} C. L\"{a}mmerzahl, \textit{The pseudodifferential operator square root of the Klein-Gordon equation}, J. Math. Phys. \textbf{34}, 3918 (1993).

\bibitem{BDQ} D. Babusci, G. Dattoli, and M. Quattromini, \textit{Relativistic equations with fractional and pseudodifferential operators}, Phys. Rev. A \textbf{83}, 062109 (2011).

\end{thebibliography}
\end{document}